\documentclass[fleqn,11pt,twoside]{article}
\usepackage[headings]{espcrc1}

\readRCS
$Id: espcrc1.tex,v 1.2 2004/02/24 11:22:11 spepping Exp $
\usepackage{graphicx}
\usepackage[figuresright]{rotating}

\newcommand{\AmS}{{\protect\the\textfont2
  A\kern-.1667em\lower.5ex\hbox{M}\kern-.125emS}}

\title{State resolved rotational excitation cross sections and rates in H$_2$+H$_2$ collisions}

\author{Renat A. Sultanov\thanks{sultanov@bcrl.stcloudstate.edu},
Dennis Guster\thanks{dguster@stcloudstate.edu}
\address{Business Computer Research Laboratory, St. Cloud State University,
2nd Floor, General Office Area, BB-252, 720 Fourth Avenue South, St Cloud, MN 56301-4498}}

\runtitle{Rotational excitations in H$_2$+H$_2$ collisions}
\runauthor{R.A. Sultanov and D. Guster}

\begin{document}

\maketitle

\begin{abstract}
Rotational transitions in molecular hydrogen collisions are computed.
The two most recently developed potential energy surfaces 
for the H$_2-$H$_2$ system are used from the following works:
1) A.I. Boothroyd, P.G. Martin, W.J. Keogh, M.J. Peterson, J. Chem. Phys., 116 (2002) 666,
and 2) P. Diep, J.K. Johnson, J. Chem. Phys., 113 (2000) 3480; ibid. 112, 4465.
Cross sections for rotational transitions 00$\rightarrow$20, 22, 40, 42, 44
and corresponding rate coefficients are calculated using a quantum-mechanical approach.
Results are compared for a wide range of kinetic temperatures 300 K $ \le T \le$ 3000 K.
\end{abstract}

\section{INTRODUCTION} 

The interaction and collision properties of hydrogen molecules, and 
hydrogen molecular isotopes has been of great theoretical and experimental interest for many years
\cite{rabitz72,farrar72,zarur74,shih75,green75,bauer76,green78,schaefer80,flower87,schwenke88,schwenke90,schaefer90,booth91,aguado94,flower98a,flower98b,billing99,flower00a,pogrebnya02,guo02,mandy04,jose05,mate05,hinde05}.
Because of the low number of electrons in the H$_2-$H$_2$ system this is one of the few four-center systems for which 
the potential energy surface (PES) can be
developed with very high precision. Therefore H$_2$+H$_2$ is a benchmark collision used for testing several
dynamic methods such as: semiclassical \cite{billing99}, quantum-mechanical \cite{pogrebnya02}
or wave packet \cite{guo02} studies.
The system may also be useful in improving our understanding of fundamental processes in few-body molecular dynamics.
Additionally, the H$_2$+H$_2$ elastic and inelastic collisions are
of interest in combustion, spacecraft modeling and in clean renewable energy.
Hydrogen gas, in particular, has generated much interest as a energy supplier, see for example \cite{zuttel04}.

The hydrogen molecule plays an important role in many areas of astrophysics 
\cite{dove87,hodapp02,shaw05,renat05}.
It is the simplest and most abundant molecule in the universe especially in giant molecular clouds. 
Energy transfer involving H$_2$ molecules governs the evolution of shock fronts \cite{dove87,hodapp02}
and photodissociation regions (PDRs) in the interstellar medium. 
Collision-induced energy transfer between H$_2$ molecules and between H$_2$ and other atoms/molecules 
is related to an important astrophysical processes, which is the cooling of
primordial gas 
and shock wave-induced heating in the interstellar media. 
To accurately model the thermal balance and kinetics of such important systems one needs accurate
state-to-state rate constants.


It is well known, that experimental measurements of quantum state resolved cross sections 
and rates is a very difficult technical problem.
On the other hand accurate theoretical data requires precise PESs and reliable dynamical treatment of the 
collision processes.
The first attempt to construct a realistic full-dimensional ab initio PES for the H$_2-$H$_2$
system was done in works \cite{schwenke88,schwenke90}, and the potential was widely used in the framework of variaty of
methods and computation techniques.

Because of the immense theoretical and practical benefits associated with the recent hydrogen fuel issues,
the H$_2$$-$H$_2$ system has been reinvestigated and
an accurate interaction potential from the first principles has been developed in work \cite{diep00}. However, in this
work the Diep and Johnson potential energy surface (DJ PES)
was extrapolated only for the rigid rotor monomer model of H$_2$$-$H$_2$.
%
On the other hand two extensive studies of the H$_2$$-$H$_2$ PES have been reported by Boothroyd et al.,
\cite{booth91,booth02}.
In these studies the potential energies have been represented at 6101 and 48180 geometries respectively
with a large basis set at the multireference configuration interaction level. 
The earlier 6101 points were fitted to a six dimensional many-body expansion form in work \cite{aguado94}.


In this work we present a comparative study of the global BMKP and DJ PESs for collisions of
rotationally excited H$_2$ molecules. The scattering cross sections and their
corresponding rate coefficients are calculated using a non reactive quantum-mechanical close-coupling
approach. In the next section we will shortly outline the method. Our results and discussion are presented in 
Section 3. Conclusions are provided in Section 4. Atomic units (e=m$_e$=$\hbar$=1) are used throughout this work.

\section{METHOD}   

In this section we will briefly present the close-coupling quantum-mechanical approach we used to calculate
our results, specifically
the cross sections and rates in collision of a hydrogen molecule with another hydrogen molecule.
The Schr\"odinger equation for a $ab+cd$ collision in the center of a mass frame,
where $ab$ and $cd$ are linear rigid rotors is
\begin{equation}
\left(\frac{P_{\vec R}}{2M_{12}}+\frac{L_{\hat r_{1}}}{2\mu_1r_1^2}+\frac{L_{\hat r_{2}}}{2\mu_2r_2^2}+
V(\vec r_1,\vec r_2,\vec R)\right)
\Psi(\hat r_{1},\hat r_{2},\vec R)=0.
\label{eq:schred}
\end{equation}
where $P_{\vec R}$ is the momentum operator of the kinetic energy of the collision, $\vec R$ is the collision 
coordinate,
$M_{12}$ is a reduced mass of the pair of two-atomic molecules (rigid rotors in this model) $ab$ and $cd$:
%
$M_{12} = (m_a+m_b)(m_c+m_d)/(m_a+m_b+m_c+m_d)$,
%
$\mu_{1(2)}$ are reduced masses of the targets:
%
$\mu_{1(2)}=m_{a(c)}m_{b(d)}/(m_{a(c)}+m_{b(d)})$,
%
$\hat r_{1(2)}$ are the angles of orientation of rotors $ab$ and $cd$, respectively, $J$ is total
angular momentum quantum number of $abcd$ system and $M$ is its projection onto the space fixed $z$ axis,
$V(\vec r_1,\vec r_2,\vec R)$ is the potential energy surface for the four atomic system $abcd$. 

The eigenfunctions of the operators $L_{\hat r_{1(2)}}$ 
in (\ref{eq:schred}) are simple spherical harmonics $Y_{j_im_i}(\hat r)$. To solve
the equation (\ref{eq:schred}) the following expansion is used \cite{green75}
\begin{equation}
\Psi(\hat r_{1},\hat r_{2},\vec R)=\sum_{JMj_1j_2j_{12}L}\frac{U^{JM}_{j_1j_2j_{12}L}(R)}{R}
\phi^{JM}_{j_1j_2j_{12}L}(\hat r_1,\hat r_2,\vec R),
\label{eq:expn}
\end{equation}
where channel expansion functions are
\begin{eqnarray}
\phi^{JM}_{j_1j_2j_{12}L}(\hat r_1,\hat r_2,\vec R) = \sum_{m_1m_2m_{12}m}C_{j_1m_1j_2m_2}^{j_{12}m_{12}}
C_{j_{12}m_{12}lm}^{JM}Y_{j_1m_1}(\hat r_1)  
Y_{j_2m_2}(\hat r_2)Y_{Lm}(\hat R),
\end{eqnarray}
here $j_1+j_2=j_{12}$, $j_{12}+L=J$, $m_1$, $m_2$, $m_{12}$ and $m$ are projections of $j_1$, $j_2$, $j_{12}$ 
and $L$ respectively.

Substitution of (\ref{eq:expn}) into (\ref{eq:schred}) provides a set of coupled second order differential equations
for the unknown radial functions $U^{JM}_{\alpha}(R)$
\begin{eqnarray}
\left(\frac{d^2}{dR^2}-\frac{L(L+1)}{R^2}+k_{\alpha}^2\right)U_{\alpha}^{JM}(R)=2M_{12}
\sum_{\alpha'} \int <\phi^{JM}_{\alpha}(\hat r_1,\hat r_2,\vec R)\nonumber \\
|V(\vec r_1,\vec r_2,\vec R)| 
\phi^{JM}_{\alpha'}(\hat r_1,\hat r_2,\vec R)>U_{\alpha'}^{JM}(R) d\hat r_1 d\hat r_2 d\hat R,
\label{eq:cpld}
\end{eqnarray}
where $\alpha \equiv (j_1j_2j_{12}L)$.
We apply the hybrid modified log-derivative-Airy propagator in the general purpose scattering program MOLSCAT 
\cite{hutson94} to solve the coupled radial equations (\ref{eq:cpld}). Additionally, we have tested other
propagator schemes included in MOLSCAT.
Our calculations revealed that other propagators can also produce quite stable results.

The log-derivative matrix is propagated to large $R$-intermolecular distances, since all experimentally observable
quantum information about the collision is contained in the asymptotic behaviour of functions 
$U^{JM}_{\alpha}(R\rightarrow\infty)$. The numerical results are matched to the known asymptotic solution to 
derive the physical scattering $S$-matrix
%
%
\begin{equation}
U_{\alpha}^J
\mathop{\mbox{\large$\sim$}}\limits_{R \rightarrow + \infty}
\delta_{\alpha \alpha'}
e^{-i(k_{\alpha \alpha}R-(l\pi/2))} 
- \left(\frac{k_{\alpha \alpha}}{k_{\alpha \alpha'}}\right)^{1/2}S^J_{\alpha \alpha'}
e^{-i(k_{\alpha \alpha'}R-(l'\pi/2))},
\end{equation}
where $k_{\alpha \alpha'}=2M_{12}(E+E_{\alpha}-E_{\alpha'})^{1/2}$ is the channel wavenumber, $E_{\alpha(\alpha')}$
are rotational channel energies and $E$ is the total energy in the $abcd$ system.
The method was used for each partial wave until a converged cross section was obtained. 
It was verified that the results are converged with respect to the number of partial waves as well as
the matching radius, $R_{max}$, for all channels included in our calculations.

Cross sections for rotational excitation and relaxation phenomena can be obtained directly from the $S$-matrix.
In particular the cross sections for excitation from $j_1j_2\rightarrow j'_1j'_2$ summed over the final $m'_1m'_2$
and averaged over the initial $m_1m_2$ are given by
\begin{eqnarray}
\sigma(j'_1,j'_2;j_1j_2,\epsilon)=\frac{\pi}{(2j_1+1)(2j_2+1)k_{\alpha\alpha'}}
\sum_{Jj_{12}j'_{12}LL'}(2J+1)|\delta_{\alpha\alpha'}- \nonumber \\
S^J(j'_1,j'_2,j'_{12}L';j_1,j_2,j_{12},L; E)|^2.
\label{eq:cross}
\end{eqnarray}
The kinetic energy is 
%
$\epsilon=E-B_1j_1(j_1+1)-B_2j_2(j_2+1)$,
%
where $B_{1(2)}$ are the rotation constants of rigid rotors $ab$ and $cd$ respectively.

The relationship between the rate coefficient $k_{j_1j_2\rightarrow j'_1j'_2}(T)$ and the corresponding
cross section $\sigma_{j_1j_2\rightarrow j'_1j'_2}(E_{kin})$ can be obtained through the following
weighted average
\begin{equation}
k_{j_1j_2\rightarrow j'_1j'_2}(T) = \frac{8k_BT}{\pi\mu}\frac{1}{(k_BT)^2}\int_{\epsilon_s}^{\infty}
\sigma_{j_1j_2\rightarrow j'_1j'_2}(\epsilon)e^{-\epsilon/k_BT}\epsilon d\epsilon,
\end{equation}
where $\epsilon = E_{total} - E_{j_1} - E_{j_2}$ is precollisional translational energy at the
translational temperature $T$ and $\epsilon_s$ is the minimum kinetic energy for the levels $j_1$ and $j_2$
to become accessible.

\section{RESULTS}   


In this section we present our results for rotational transitions in collisions between 
$para/para$-hydrogen molecules:
\begin{equation}
\mbox{H}_2(j_1=0) +\mbox{H}_2(j_2=0) \rightarrow \mbox{H}_2(j'_1) + \mbox{H}_2(j'_2).
\label{eq:h2h2}
\end{equation}
We apply the newest PESs from the works \cite{diep00} and \cite{booth02}. The first one, DJ PES, is constructed for
the vibrationally averaged rigid monomer model of the H$_2$$-$H$_2$ system to the complete basis set limit using 
coupled-cluster theory with single, double and triple excitations. A four term spherical harmonics expansion 
model was chosen to fit the surface. It was demonstrated, that
the calculated PES can reproduce the quadrupole moment to within 0.58 \% and 
the experimental well depth to within 1 \%.

The bond length was fixed at 1.449 a.u. or 0.7668 \r{A}. DJ PES is 
defined by the center-of-mass intermolecular distance, $R$, and three angles: $\theta_1$ and $\theta_2$ are the 
plane angles and $\phi_{12}$ is the relative torsional angle. The angular increment for each of the three angles 
defining the
relative orientation of the dimers was chosen to be $30^{\circ}$. There are 37 unique configurations for each radial
separation when the symmetry of the H$_2$$-$H$_2$ system is considered. In previous works calculating the
potential a much smaller set of angular configurations designed to represent the full surface was used.
The potential was calculated from only 2.0 to 10.0 \r{A} of
intermolecular (center-of-mass) separation with an increment 0.2 \r{A}. However, near the potential minimum which is
from
2.7 to 4.5 \r{A} the grid spacing is 0.1 \r{A}. The functional form of the potential represents an expansion on
Legendre polynomials \cite{green75}.

The second potential, BMKP PES, is a global six-dimensional potential energy surface for two hydrogen molecules.
It was especially constructed to represent the whole interaction region of the chemical reaction dynamics of the 
four-atomic system and to provide an accurate as possible the van der Waals well. The ground state 
and a few excited-state energies were calculated. The new potential fits the van der Waals well to an accuracy 
within about 5\% and has an rms error of 1.43 millihartree relative to the 48180 ab initio energies. 
For the 39064 ab initio energies that lie 
below twice the H$_2$ dissociation energy BMKP PES has an rms error of 0.95 millihartree. These rms errors are
comparable to the estimated error in the ab initio energies themselves.
In the six-dimensional conformation space of the four atomic system the conical intersection forms a complicated 
three-dimensional hypersurface. The authors of the work \cite{booth02} mapped out a large portion of the locus 
of this conical intersection.

The BMKP PES uses cartesian coordinates to compute distances between 
four atoms. We have devised some fortran code, which converts spherical coordinates used in
Sec. 2 to the corresponding cartesian coordinates and computes the distances between the four atoms. In all our
calculations with this potential the bond length was fixed at 1.449 a.u. or 0.7668 \r{A} as in DJ PES.

Now we will present our results for the elastic and inelastic integral cross sections and rate coefficients for the
collision 
(\ref{eq:h2h2}). As far as astrophysical applications are concerned, we are particularly interested in the pure \
rotational transitions of the H$_2$ molecules.

A large number of test calculations have been done to secure the convergence of the results with respect to all 
parameters that enter into the propagation of the Schr\"odinger equation (\ref{eq:schred}). 
This includes the intermolecular distance $R$, the total angular momentum $J$ of the four atomic system, $N_{lvl}$ 
the number of rotational levels to be included in the close coupling expansion and others
(see the MOLSCAT manual \cite{hutson94}).

We reached convergence for the integral cross sections, $\sigma(E_{kin})$, in all considered collisions. In the case 
of DJ PES the propagation has been done from 2 \r{A} to 10 \r{A}, since this potential is defined only for the 
specific distances. For the BMKP PES we used $r_{min}=1$ \r{A} to $r_{max}=30$ \r{A}. We also applied a few different 
propagators included in the MOLSCAT program. 

Table 1 represents the convergence test results with respect to $J_{max}$, the maximum value of the total angular
momentum, for both the BMKP and DJ PESs. The calculations are limited to just three values of energy, 
for the simpler basis set: $j_1j_2$=00, 20 and 22, and
from lowest to highest within the considered range of energies.
As can be seen the results are stable for the range of kinetics energies, when $J_{max}$ is 
increased from 80 to 90. In all our subsequent production calculations we use $J_{max}=80$.

It is important to point out
here, that for comparison purposes we don't include the compensating factor of 2 mentioned in \cite{flower87}.
However, in Fig.\ 2 and in our subsequent calculations of the rate coefficients, $k_{jj'}(T)$,
the factor is included.


In Table 2 we include the results of our test calculations for the various rotational 
levels $j_1j_2$ included in the close coupling expansion.
%
In these test calculations we used two basis sets: $j_1j_2$=00, 20, 22, 40, 42 with total
basis set size $N_{lvl}=13$ and $j_1j_2$=00, 20, 22, 40, 42, 44, 60, 62 with $N_{lvl}=28$.
One can see that the
results are quite stable for the 00$\rightarrow$20 and 00$\rightarrow$22 transitions and somewhat
stable for the highly excited 00$\rightarrow$40 transition. Nontheless, for our production calculations we
used the first basis set.


The objective of this work is to make reliable quantum-mechanical calculations for different transitions
in $p$-H$_2$+$p$-H$_2$ collisions
and provide a comparative study of the two PESs. The energy dependence of the state-resolved integral cross sections
$\sigma_{j_1j_2\rightarrow j'_1j'_2}(E_{kin})$
for the $j_1=j_2=0 \rightarrow j'_1=2,j'_2=0$ and $j_1=j_2=0 \rightarrow j'_1=2,j'_2=2$ rotational
transitions are represented in Fig.\ 1 (upper plots)
for both the BMKP and DJ PESs respectively. These channels for the most part influence the total cross-section
of the H$_2$+H$_2$ collision. As can be seen both PESs provide the
same type of the behaviour in the cross section. These results are in basic agreement with the recent calculations 
of work \cite{guo02}, where the BMKP PES was also applied, 
but using a time-dependent quantum-mechanical approach.
Our calculation show, that DJ PES generates higher values for the cross sections, by up to 50\%.

\begin{sidewaystable}  
\caption{Convergence of the total cross sections $(10^{-16}\mbox{cm}^2)$ for transitions
00$\rightarrow$20 and 00$\rightarrow$22 with respect to the maximum value of the total angular
momentum $J_{max}$ in the H$_2-$H$_2$ system. Here $\sigma_{\mbox{\tiny{B}}}$ and $\sigma_{\mbox{\tiny{D}}}$
are the cross sections calculated with BMKP \cite{booth02} and Diep and Johnson \cite{diep00} PESs respectively.}
\label{table:1}
\begin{tabular*}{\textheight}{@{\extracolsep{\fill}}ccccccccccccc}
\hline
&\multicolumn{4}{c}{$J_{max}=80$} & \multicolumn{4}{c}{$J_{max}=90$}\\
&\multicolumn{2}{c}{00$\rightarrow$20} & \multicolumn{2}{c}{00$\rightarrow$22} &
 \multicolumn{2}{c}{00$\rightarrow$20} & \multicolumn{2}{c}{00$\rightarrow$22} \\
\hline
$E$ (eV) 
& $\sigma_{\mbox{\tiny{B}}}$ & $\sigma_{\mbox{\tiny{D}}}$ & 
  $\sigma_{\mbox{\tiny{B}}}$ & $\sigma_{\mbox{\tiny{D}}}$ & 
  $\sigma_{\mbox{\tiny{B}}}$ & $\sigma_{\mbox{\tiny{D}}}$ & 
  $\sigma_{\mbox{\tiny{B}}}$ & $\sigma_{\mbox{\tiny{D}}}$ \\
\hline
  1.240 & 2.410     &  3.127      &  2.549     &  3.925     & 2.410     &  3.127     & 2.552     &  3.933    \\
  0.620 & 1.910     &  2.485      &  1.146     &  1.908     & 1.910     &  2.485     & 1.146     &  1.909    \\
  0.124 & 1.742(-1) &  5.403(-1)  &  1.526(-2) &  2.312(-2) & 1.742(-1) &  5.403(-1) & 1.526(-2) &  2.312(-2)\\
%
%
\hline
\end{tabular*}\\[2pt]
Numbers in parentheses are powers of 10 (the compensating factor of 2 is not included).

\vspace{15mm}

\caption{Convergence of the total cross sections $(10^{-16}\mbox{cm}^2)$ for transitions from
00$\rightarrow$20, 22, 40 with respect to the number $N_{lvl}$ of the levels to be
included in the basis set of the H$_2-$H$_2$ system. Here $\sigma_{\mbox{\tiny{B}}}$ and $\sigma_{\mbox{\tiny{D}}}$
are the cross sections calculated with BMKP \cite{booth02} and Diep and Johnson \cite{diep00} PESs respectively.}
\label{table:2}
\renewcommand{\arraystretch}{1.2} 
\begin{tabular*}{\textheight}{@{\extracolsep{\fill}}ccccccccccccc}
\hline
%
\hline
&\multicolumn{6}{c}{Basis set: $j_1j_2$=00, 20, 22, 40, 42 ($N_{lvl}=13$)}
&\multicolumn{6}{c}{Basis set: $j_1j_2$=00, 20, 22, 40, 42, 44, 60, 62 ($N_{lvl}=28$)}\\
&\multicolumn{2}{c}{00$\rightarrow$20} & \multicolumn{2}{c}{00$\rightarrow$22} & \multicolumn{2}{c}{00$\rightarrow$40}
&\multicolumn{2}{c}{00$\rightarrow$20} & \multicolumn{2}{c}{00$\rightarrow$22} & \multicolumn{2}{c}{00$\rightarrow$40}\\
\hline
$E$ (eV) & $\sigma_{\mbox{\tiny{B}}}$ & $\sigma_{\mbox{\tiny{D}}}$ & 
           $\sigma_{\mbox{\tiny{B}}}$ & $\sigma_{\mbox{\tiny{D}}}$ & 
           $\sigma_{\mbox{\tiny{B}}}$ & $\sigma_{\mbox{\tiny{D}}}$ & 
           $\sigma_{\mbox{\tiny{B}}}$ & $\sigma_{\mbox{\tiny{D}}}$ & 
           $\sigma_{\mbox{\tiny{B}}}$ & $\sigma_{\mbox{\tiny{D}}}$ & 
           $\sigma_{\mbox{\tiny{B}}}$ & $\sigma_{\mbox{\tiny{D}}}$ \\
\hline
1.240 & 2.50     & 3.14    & 1.97    & 3.00    & 7.32(-2) & 1.18(-1) & 2.48    &3.13    &1.99    &2.97    &8.07(-2)&1.26(-1)\\
0.620 & 1.94     & 2.55    & 1.07    & 1.73    & 2.72(-2) & 4.38(-2) & 1.93    &2.55    &1.07    &1.73    &2.80(-2)&4.59(-2)\\
0.124 & 1.75(-1) & 5.45(-1)& 1.54(-2)& 2.32(-2)& 0.0       & 0.0     & 1.75(-1)&5.44(-1)&1.54(-2)&2.32(-2)&0.0     &0.0     \\
%
%
%
\hline
\end{tabular*}\\[2pt]
Numbers in parentheses are powers of 10 (the compensating factor of 2 is not included).
\end{sidewaystable}  

\begin{figure}
\begin{center}
\includegraphics*[width=25pc,height=18pc]{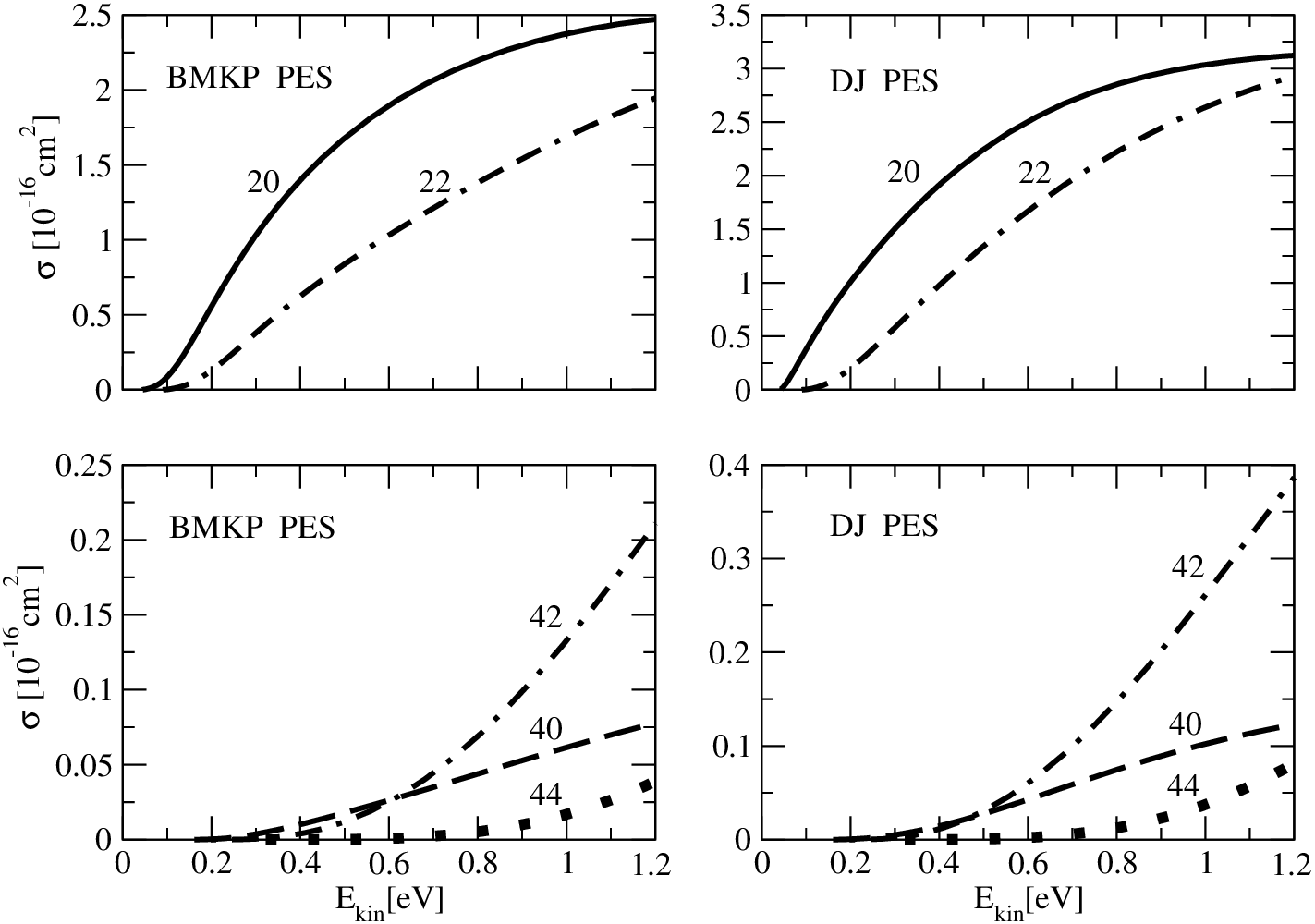}  
\end{center}
\caption{Rotational state resolved integral cross sections for $j_1=j_2=0 \rightarrow j'_1=2,j'_2=0$,
$j_1=j_2=0 \rightarrow j'_1=2,j'_2=2$, $j_1=j_2=0 \rightarrow j'_1=4,j'_2=0$,
$j_1=j_2=0 \rightarrow j'_1=4,j'_2=2$ and $j_1=j_2=0 \rightarrow j'_1=4,j'_2=4$ calculated with the BMKP and DJ
PESs (the compensating factor of 2 is not included).}
\label{fig:fig1}
\end{figure}
\begin{figure}
\begin{center}
\includegraphics[scale=1.0,width=23pc,height=17pc]{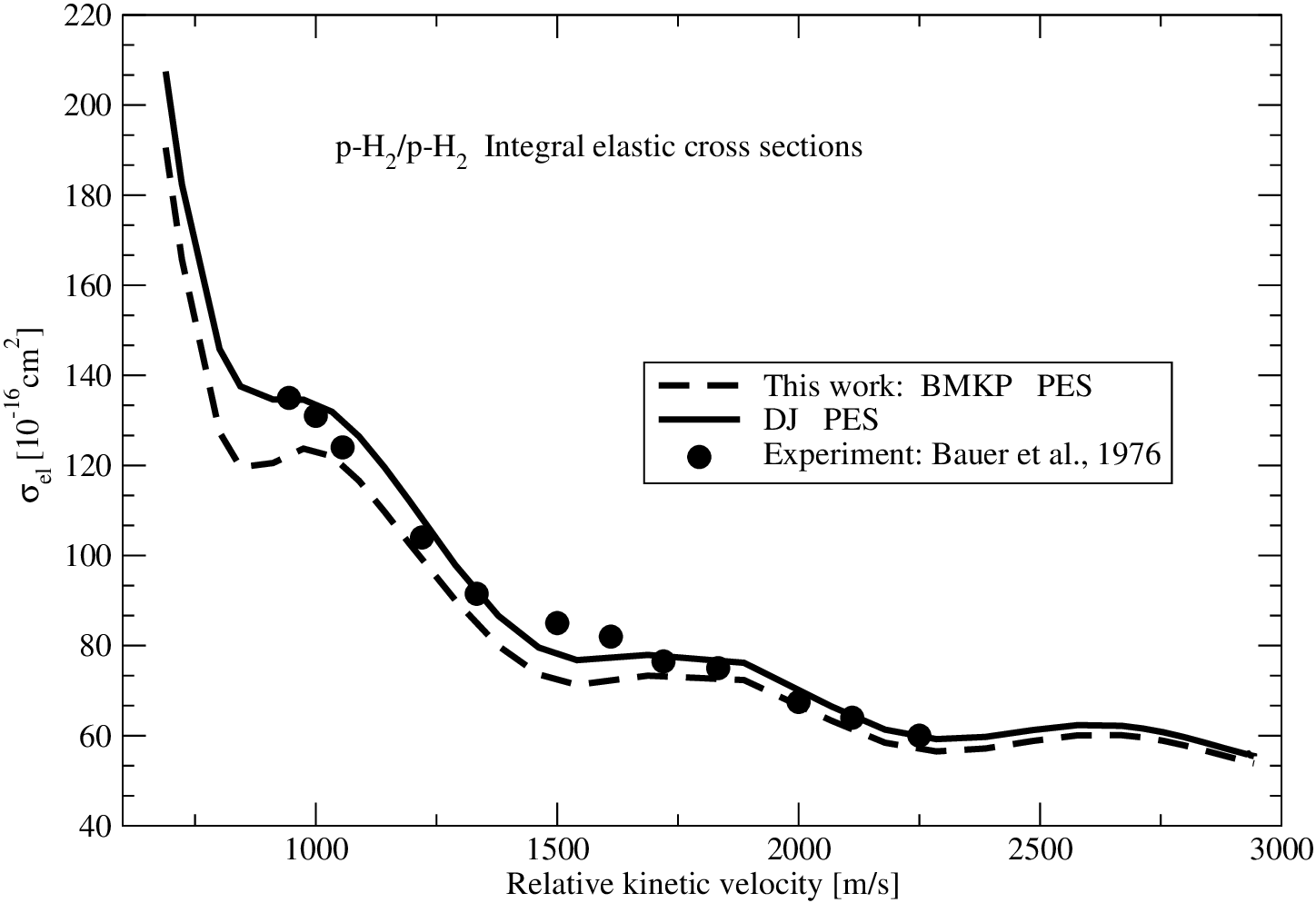}  
\end{center}
\caption{Integral elastic cross sections calculated with the BMKP and DJ potentials. The experimental
measurements are those of Bauer and co-workers \cite{bauer76} (the compensating factor of 2 is included).}
\label{fig:fig2}
\end{figure}

\begin{figure}
\begin{center}
\includegraphics[scale=1.0,width=23pc,height=17pc]{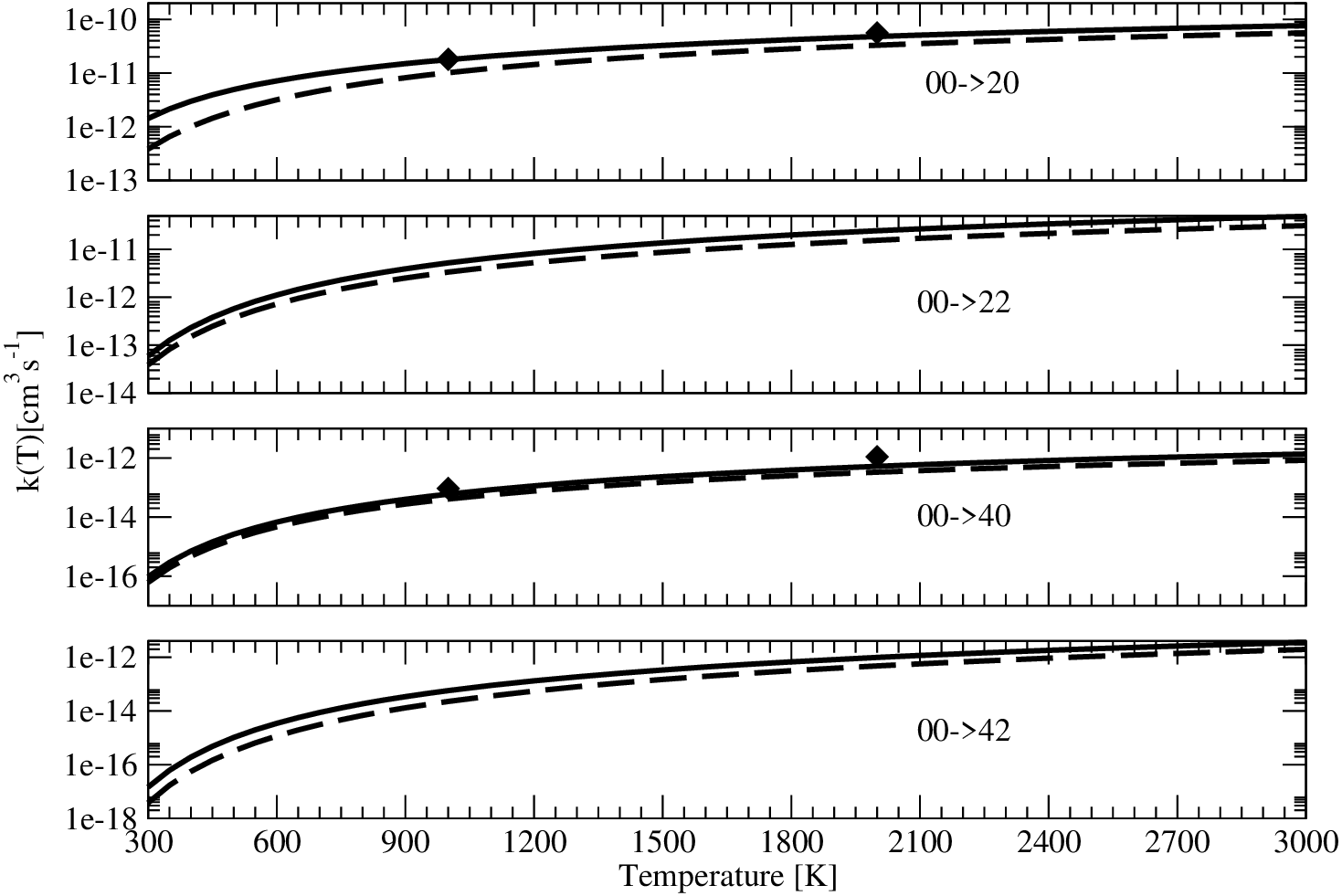} 
\end{center}
\caption{Temperature dependence of the state-resolved thermal rate constants for the
$j_1=j_2=0 \rightarrow j'_1=2,j'_2=0, j'_1=2,j'_2=2,  j'_1=4,j'_2=0$ and $j'_1=4,j'_2=2$.
The results for the DJ and BMKP PESs are given in solid and dashed lines, respectively.
The diamonds are the theoretical data of work \cite{flower98b}.}
\label{fig:fig3}
\end{figure}

\begin{table}[htb]
\caption{Rate coefficients $k_{00\rightarrow jj'}(T)$ (cm$^3$s$^{-1}$) calculated with the DJ PES
for rotational transitions $00\rightarrow 02, 22$ and 40 in comparison with other theoretical \cite{flower98a} and 
experimental \cite{mate05} data.}
\label{table:3}
\begin{tabular}{lcccccc}
\hline
&\multicolumn{3}{c}{$k_{00\rightarrow 20}$} & \multicolumn{2}{c}{$k_{00\rightarrow 22}$} 
&$k_{00\rightarrow 40}$\\
\hline
$T (K)$ & This work & \cite{flower98a} & \cite{mate05} & This work & \cite{flower98a} & This work \\
\hline\hline
     50   &   6.25(-17) &          & 1.1$\pm$0.1 (-16) &    1.71(-22)  &           &            \\
     60   &   3.80(-16) & 4.4(-16) & 6.0$\pm$0.7 (-16) &    6.42(-21)  & 6.9(-21)  &            \\
     100  &   1.64(-14) & 1.6(-14) & 2.2$\pm$0.4 (-14) &    1.13(-17)  & 0.97(-17) &   1.20(-22)\\
     110  &   2.82(-14) &          & 3.6$\pm$0.6 (-14) &    3.26(-17)  &           &   6.89(-22)\\
     120  &   4.47(-14) &          &                   &    8.02(-17)  &           &   3.01(-21)\\
     150  &   1.28(-13) & 1.2(-13) &                   &    6.16(-16)  & 4.5(-16)  &   8.14(-20)\\
     200  &   3.98(-13) &          &                   &    5.37(-15)  &           &   2.48(-18)\\
     300  &   1.43(-12) &          &                   &    5.94(-14)  &           &   9.59(-17)\\
\hline
\end{tabular}\\[2pt]
Numbers in parentheses are powers of 10.
\end{table}

The integral cross sections for inelastic collisions:
$j_1=j_2=0 \rightarrow j'_1=4,j'_2=0$, $j_1=j_2=0 \rightarrow j'_1=4,j'_2=2$, 
and $j_1=j_2=0 \rightarrow j'_1=4,j'_2=4$ for BMKP and DJ PESs are presented in Fig.\ 1 (bottom plots).
The cross sections are
very small at energies less than 0.25 eV, since at lower kinetic energies these transitions are closed by 
the corresponding energy barriers.

We would like to point out here, that the $j_1=j_2=0 \rightarrow j'_1=4,j'_2=2$ cross section
becomes larger than $j_1=j_2=0 \rightarrow j'_1=4,j'_2=0$ if the collision energy is greater than $\sim$0.6 eV.
In this energy range it is hence more likely that the second diatom is also excited when the first diatom
makes the $0 \rightarrow 4$ transition. However the $j_1=0,j_2=0 \rightarrow j'_1=4,j'_2=4$ cross section is very small
over the entire energy range considered.
The DJ potential energy surface again provides higher
results in the corresponding cross sections. One can note, that for both potentials 
the rotational inelasticity is dominated by the $00\rightarrow 20$ transition.

The integral cross sections for an elastic $p$-H$_2$+$p$-H$_2$ collision computed 
with both PESs are depicted in Fig.\ 2. The two cross sections are in reasonable agreement 
over a wide range of energies. Again the BMKP PES generates higher values. The difference becomes even larger
at lower kinetic energies. In the figure we also provide experimental data from work \cite{bauer76}.
The theoretical and the experimental results are in reasonable agreement with each other. This fact indicates,
that the spherical parts of BMKP and DJ PESs are close in shape, which is a very important attribute for PESs.

The differences in the cross sections between the two potentials are also reflected in the state-resolved
thermal rate constants, as shown in Fig.\ 3. Again, the BMKP PES underestimates the rate constant for the
00$\rightarrow$ 20 transition, and overestimates those transitions of higher rotational levels. The near
perfect agreement for the 00$\rightarrow$ 40 transition is likely accidental. Because of the highly averaged
nature of the rate constant, the difference is not as conspicuous as in the cross sections. We have also plotted
in the same figure the rate constants reported previously in work \cite{flower98a}, where Schwenke's PES
\cite{schwenke88} was used.

Finally, our rate constants for 
00 $\rightarrow$ 20, 22, 40 rotational transitions calculated with the DJ PES at lower temperatures are listed in
Table 3 together with the theoretical calculations of Flower et al. \cite{flower98a} 
and recent experimental results from work \cite{mate05}. As can be seen our rate constants $k_{00\rightarrow 20}(T)$ 
and $k_{00\rightarrow 22}(T)$ are close to those of work \cite{flower98a}. The experimental results are higher by
about 60\% for 50 K and 30 \% for 110 K.


\section{SUMMARY and CONCLUSIONS}     

A systematical study of the state-resolved rotational excitation cross sections and rates
in molecular $para$-/$para$-hydrogen collisions 
is completed. A test of convergence and the results for cross sections and rate coefficients
using two different potential energy surfaces for the H$_2-$H$_2$
system have been obtained for a wide range of kinetic energies.
%

Although our calculations revealed, that both PESs can provide the
same type of behaviour in regard to cross sections 
and rates, there are still significant differences. The DJ potential overestimates by
about 20-40 \% the
results at even relatively larger kinetic energies. This is especially true in regard to 00$\rightarrow$20 rotational
transition, where significant differences at around 300 K are seen in Fig. 3.

Considering the results of these calculations one can conclude that subsequent work is needed to further improve the 
H$_2-$H$_2$ PES. Detailed calculations including rotational-vibrational basis set
and comparative analyses using both potentials at low and very low kinetic energies 
for $o$-H$_2$/$o$-H$_2$ and $p$-H$_2$/$o$-H$_2$ excitation-deexcitation collision processes currently are in 
progress in our group.

\vspace{5mm}

\noindent{{\bf Acknowledgments}}


This work was supported by the St. Cloud State University internal grant program, St. Cloud, MN (USA).


\begin{thebibliography}{99}

\bibitem{rabitz72} H. Rabitz, J. Chem. Phys., 57, (1972) 1718.

\bibitem{farrar72} J.M. Farrar, Y.T. Lee, J. Chem. Phys., 57 (1972) 5492.

\bibitem{zarur74} G. Zarur, H. Rabitz, J. Chem. Phys., 60 (1974) 2057

\bibitem{shih75} S.-I. Chu, J. Chem. Phys., 62, (1975) 4089.

\bibitem{green75} S. Green, J. Chem. Phys., 62 (1975) 2271; J. Chem. Phys., 67 (1977) 715.

\bibitem{bauer76} W. Bauer, B. Lantzsch, J.P. Toennies, K. Walaschewski, Chem. Phys. 17 (1976) 19.


\bibitem{green78} T.G. Heil, S. Green, D.J. Kouri, J. Chem. Phys., 68 (1978) 2562.

\bibitem{schaefer80} L. Monchick, J. Schaefer, J. Chem. Phys., 73 (1980) 6153.

\bibitem{flower87} G. Danby, D.R. Flower, T.S. Monteiro, Mon. Not. R. Astr. Soc., 226 (1987) 739.

\bibitem{schwenke88} D.W. Schwenke, J. Chem. Phys., 89 (1988) 2076.  

\bibitem{schwenke90} D.W. Schwenke, J. Chem. Phys., 92 (1990) 7267.

\bibitem{schaefer90} J. Schaefer, Astron. Astrophys. Suppl. Ser., 85 (1990) 1101.

\bibitem{booth91} A.I. Boothroyd, W.J Keogh, P.G. Martin, M.J. Peterson, J. Chem. Phys., 95 (1991) 4331.


\bibitem{aguado94} A. Aguado, C. Suarez, M. Paniagua, J. Chem. Phys., 101 (1994) 4004.



\bibitem{flower98a} D.R. Flower, Mon. Not. R. Astron. Soc., 297 (1998) 334.

\bibitem{flower98b} D.R. Flower, E. Roueff, J. Phys. B: At. Mol. Opt. Phys., 31 (1998) 2935.


\bibitem{billing99} V.A. Zenevich, G. D. Billing, J. Chem. Phys., 111 (1999) 2401.

\bibitem{flower00a} D.R. Flower, J. Phys. B: At. Mol. Opt. Phys., 33 (2000) L193.



\bibitem{pogrebnya02} S.K. Pogrebnya, D.C. Clary, Chem. Phys. Lett., 363 (2002) 523.

\bibitem{guo02} S.Y. Lin, H. Guo, J. Chem. Phys., 117 (2002) 5183.


\bibitem{mandy04} M.E. Mandy, S.K. Pogrebnya, J. Chem. Phys., 120 (2004) 5585.

\bibitem{jose05} M. Bartolomei, M.I. Hernandez, J. Campos-Martinez, J. Chem. Phys., 122 (2005) 064305.

\bibitem{mate05} B. Mate, F. Thibault, G. Tejeda, J.M. Fernandez, S. Montero, J. Chem. Phys., 122 (2005) 064313.

\bibitem{hinde05} R.J. Hinde, J. Chem. Phys., 122 (2005) 144304.

\bibitem{zuttel04} A. Z\"uttel, Naturwissenschaften, 91 (2004) 157.








\bibitem{dove87} J.E. Dove, A.C.M. Rusk, P.H. Cribb, P.G. Martin, Astrophys. J., 318 (1987) 379.




\bibitem{hodapp02} K.W. Hodapp, C.J. Davis, Astrophys. J., 575 (2002) 291.


\bibitem{shaw05} G. Shaw, G.J. Ferland, N.P. Abel, P.C. Stancil, P.A.M. van Hoof, Astrophys. J. 624 (2005) 794.

\bibitem{renat05} R.A. Sultanov, N. Balakrishnan, Astrophys. J. 629 (2005) 305.




\bibitem{diep00} P. Diep, J.K. Johnson, J. Chem. Phys., 113 (2000) 3480; ibid. 112 (2000) 4465.

\bibitem{booth02}A.I. Boothroyd, P.G. Martin, W.J. Keogh, M.J. Peterson, J. Chem. Phys., 116 (2002) 666.

\bibitem{hutson94} J.M. Hutson, S. Green, MOLSCAT VER. 14 (1994)
(Distributed by Collabor. Comp. Proj. 6, Daresbury Lab., UK, Eng. Phys. Sci. Res. Council, 1994)
\end{thebibliography}
\end{document}